%% file: main.tex
\documentclass[%
 reprint,
superscriptaddress,
 amsmath,amssymb,
 aps,
 longbibliography, 
prb,
]{revtex4-2}

\usepackage{graphicx}
\usepackage{dcolumn}
\usepackage{bm}
\usepackage{amsmath}
\usepackage{blkarray}
\usepackage{color}
\usepackage{url}
\usepackage[colorlinks=true,breaklinks=true]{hyperref}
\usepackage{float}
\usepackage{multirow}
\usepackage{makecell}

\newcommand\identity{1\kern-0.25em\text{l}}

\begin{document}

\title{Enhancing the Expressivity of Variational Neural, and Hardware-Efficient Quantum States Through Orbital Rotations}

\author{Javier Robledo Moreno}
\email{jrm874@nyu.edu}
\affiliation{Center for Computational Quantum Physics, Flatiron Institute, New York, New York, 10010, USA}
\affiliation{Center for Quantum Phenomena, Department of Physics, New York University, 726 Broadway, New York, New York 10003, USA}
\affiliation{IBM Quantum, IBM Research - 1101 Kitchawan Rd, Yorktown Heights, New York 10598, USA}
\author{Jeffrey Cohn}
\affiliation{IBM Quantum, IBM Research - Almaden, 650 Harry Road, San Jose, CA 95120, USA} 
\author{Dries Sels}
\affiliation{Center for Computational Quantum Physics, Flatiron Institute, New York, New York, 10010, USA}
\affiliation{Center for Quantum Phenomena, Department of Physics, New York University, 726 Broadway, New York, New York 10003, USA}
\author{Mario Motta}
\affiliation{IBM Quantum, IBM Research - Almaden, 650 Harry Road, San Jose, CA 95120, USA}


\date{\today}

\begin{abstract}
Variational approaches, such as variational Monte Carlo (VMC) or the variational quantum eigensolver (VQE), are powerful techniques to tackle the ground-state many-electron problem. Often, the family of variational states is not invariant under the reparametrization of the Hamiltonian by single-particle basis transformations. As a consequence, the representability of the ground-state wave function by the variational {\it ansatz} strongly dependents on the choice of the single-particle basis. In this manuscript we study the joint optimization of the single-particle basis, together with the variational state in the VMC (with neural quantum states) and VQE (with hardware-efficient circuits) approaches. We show that the joint optimization of the single-particle basis with the variational state parameters yields significant improvements in the expressive power and optimization landscape in a variety of chemistry and condensed matter systems. 
We also realize the first active-space calculation using neural quantum states, where the single-particle basis transformations are applied to all of the orbitals in the basis set.
\end{abstract}

\maketitle


\section{Introduction}

The description of the ground-state properties of an interacting electron system suffers from the difficulty that any wave function belongs to an exponentially-scaling Hilbert space in the number of single-particle basis elements (orbitals) that defines the system. The Hartree-Fock method is an approximate technique with polynomial scaling based on single-particle orbital rotations. It produces uncorrelated wave functions (Slater determinants), capable of qualitatively describing a variety of ground-state phenomena~\cite{Zhang2008HF, Xu2011HF, Schultz1990HF, Poilblanc1989HF}. However, subtle many-body effects, like those present in reaction pathways and transitions states~\cite{krylov1998EthyleneCusps} or Mott physics~\cite{Mott1949} amongst others, require linear combinations of Slater determinants to accurately describe the ground-state properties of the system.

Variational techniques propose a trial wave function defined by a set of variational parameters, which are determined by the optimization of the expectation value of the Hamiltonian. Some examples of trial wave functions include quantum circuits (variational quantum eigensolver~\cite{Peruzzo2014OG-VQE}), weighted and parametrized linear combinations of all possible Slater determinants where the relevant terms are sampled according to Monte-Carlo techniques (variational Monte Carlo~\cite{McMillan1965OG-VMC}), weighted linear combinations of small restricted sub-sets of Slater determinants (configuration interaction (CI) and its extension to active spaces (CAS-CI)~\cite{roos2005multiconfigurational}), or tensor networks optimized using the density matrix renormalization group (DMRG)~\cite{White1992OG-DMRG, White1993OG-DMRG}. All of these approaches have in common the ability to represent a ground-state wave function given by linear combinations of Slater-determinants. 
\begin{figure}[H]
\centering
    \includegraphics[width = .75 \columnwidth]{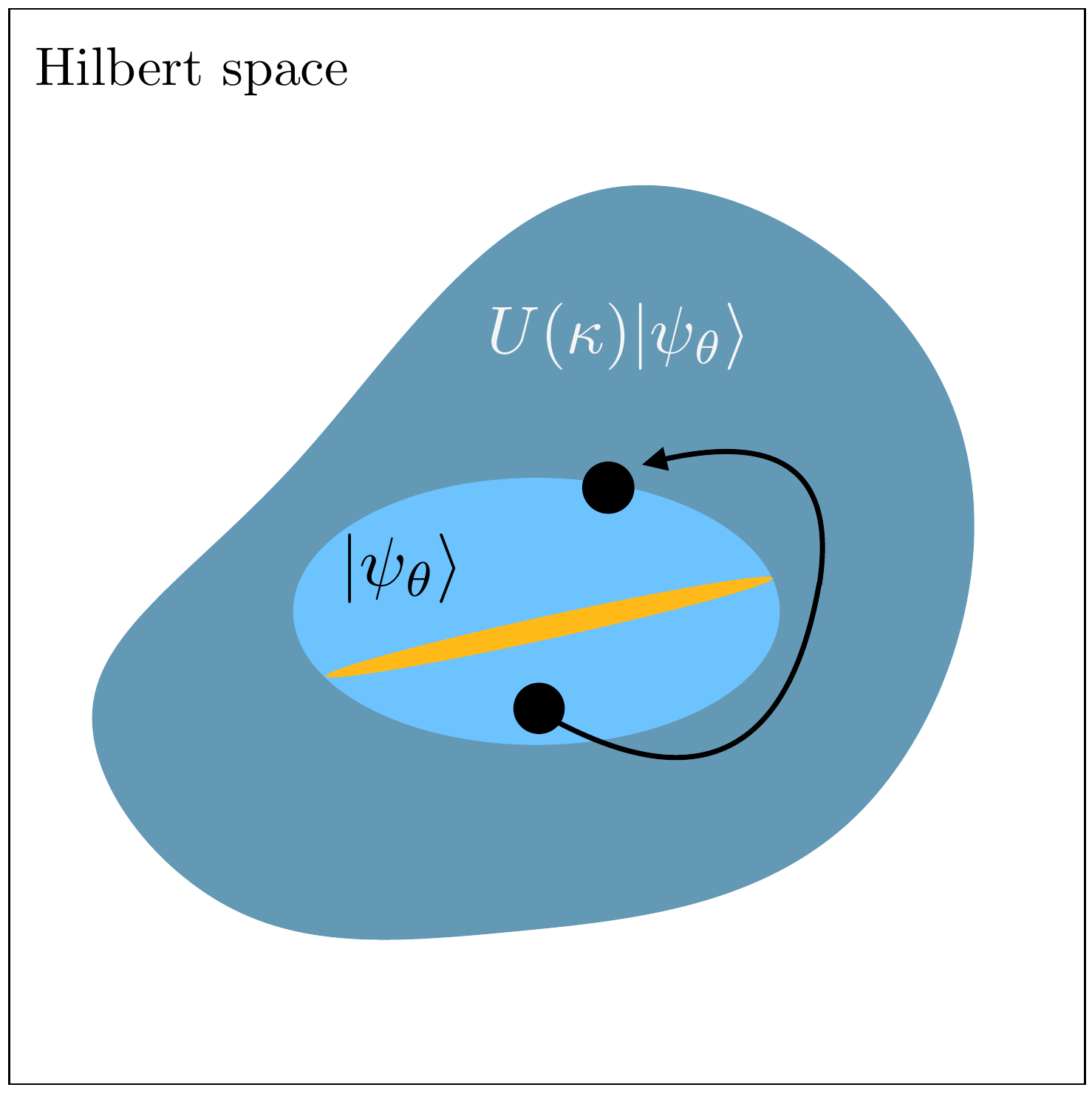}
    \caption{\label{fig_00}  Schematic depiction of the variational manifold. By dressing the variational state $|\psi_\theta\rangle$ with single-particle orbital rotations, i.e. apply the parametrized Gaussian transformation $U(\kappa)$ (see Eq.~\ref{Eq04:rotation parametrization}) to $|\psi_\theta\rangle$, one can significantly increase its expressive power and optimization landscape. The light blue ellipse represents the set of states accessible to $|\psi_\theta\rangle$, and the dark blue shape represents the set of states accessible to the dressed variational state $U(\kappa)|\psi_\theta \rangle$ (see Eq.~\ref{Eq08:variationalRotated}). In a fixed single-particle basis, some states (black dots) are not connected via the optimization procedure due to barriers in the optimization landscape, represented by the orange line. Including orbital rotations can provide alternative paths (black arrow) to connect those states, while covering a larger region of the Hilbert space.}
\end{figure}
\noindent However, the above post-Hartree-Fock methods are not invariant under the reparametrization of the Hamiltonian defined by orbital rotations. As a consequence, the representability of the ground state by the trial wave function is affected by the choice of the single-particle basis that describes the problem.

For the study of molecular systems, the set of orthogonal Hartree-Fock orbitals, also known as molecular orbitals (MOs), is the canonical choice of a basis to construct linear combinations of Slater determinants as the variational {\it ansatz}. This choice is motivated by the observation that the Hartree-Fock determinant is the dominant wave function amplitude in the ground state. However, it was observed that active-space methods on a fixed basis yield nonphysical results, like potential energy surfaces with cusps~\cite{krylov1998EthyleneCusps}. In order to eliminate the undesirable effects of working on a fixed basis, self consistent field (SCF) approaches were proposed in the context of active space calculations (CAS-SCF), where the choice of the single-particle basis becomes part of the variational procedure~\cite{roos1980complete, head1988optimization, werner1985second, olsen2011casscf}. The optimization of orbital rotations and variational-{\it ansatz} parameters is carried out in alternating order until convergence is reached. Orbital optimizations have been generalized to operate within restricted active spaces~\cite{malmqvist1990restricted} and  DMRG~\cite{zgid_density_2008, ghosh_orbital_2008, wouters_density_2014, ma_second-order_2017, wu_disentangling_2022} in the interest of including more orbitals in the active space.  More recently, orbital optimizations have been combined with VQE approaches~\cite{sokolov_quantum_2020, mizukami_orbital_2020, bierman_improving_2022, yalouz_state-averaged_2021, yalouz_analytical_2022, omiya_analytical_2022, tao_analytical_2021, nakagawa_analytical_2022}. 

In this paper we investigate the effect of variational orbital rotations in the context of VMC using neural quantum states (NQS)~\cite{Carleo2017Science,hermann2022ReviewPaperNQS} as the variational {\it ansatz}. The parameters that control the orbital rotations are treated on equal footing as the variational parameters of the NQS, and are optimized simultaneously using gradient descent techniques. We show that the addition of parametrized orbital rotations to the variational state allows to adapt, \textit{on the fly},  the single-particle basis where the amplitudes are defined. This optimization procedure yields the single-particle basis where the expressive power of the variational state is maximized to represent the ground state. In some cases, the single-particle basis rotations yield ground-state wave functions whose amplitudes are positive semidefinite or that define a single-mode probability distribution, which are therefore easier to sample from. We provide numerical evidence showing that this joint optimization improves the accuracy of the variational trial state on a variety of many-body problems including \emph{ab initio} Hamiltonians and lattice models. Different classes of NQS are considered, including determinant-based and non-determinant based states. 
 Furthermore, we demonstrate, for the first time, an active-space variational calculation using NQS, in which the active-space orbitals and NQS parameters are are variationally optimized to self-consistency. We observe significant improvements when compared to a calculation employing MOs as active-space orbitals, therefore enabling more accurate variational calculations than otherwise possible.
We also investigate the improvement of the variational power of hardware-efficient quantum circuits in VQE. In that case, orbital optimizations can be treated as a classical post-processing step which does not increase the variational circuit depth. We find that the accuracy on the variational-ground state energy is considerably increased by the addition of the orbital rotations. We also study the characteristics of the optimization dynamics, and find that orbital rotations improve the energy landscape in the space of variational parameters.  Figure~\ref{fig_00} provides an schematic depiction of the improvement of the expressivity and optimization landscape of variational quantum states via orbital rotations.

\section{Methods}

Consider the interacting-electron Hamiltonian in second quantization projected into a basis set with $N_{\textrm{orb}}$ basis elements,
\begin{equation} \label{Eq01:Hmailtonian}
    \begin{split}
    \hat{H} = &  \sum_{\substack{pq \\ \sigma}} h_{pq} \hat{c}^\dagger_{p\sigma} \hat{c}_{q\sigma} + \\
    & \sum_{\substack{pqrs \\ \sigma}}  \frac{h_{pqrs}}{2}\left( \hat{c}^\dagger_{p\sigma} \hat{c}^\dagger_{q\sigma} \hat{c}_{s\sigma}  \hat{c}_{r\sigma} +
     \hat{c}^\dagger_{p\sigma} \hat{c}^\dagger_{q -\sigma} \hat{c}_{s -\sigma}  \hat{c}_{r\sigma}\right)
    \end{split}
\end{equation}
where $1\leq p, q, r, s \leq N_{\textrm{orb}}$ are the orbital labels and $\sigma \in \{1, -1\}$ labels the spin degree of freedom. $h_{pq}$ and $h_{pqrs}$ are the one- and two-body integrals for the single-particle orbital functions that define the basis. The number of both spin-up and spin-down electrons is fixed. The anti-commuting algebra of the fermionic operators is encoded via the Jordan-Wigner mapping~\cite{JordanWigner1928}.

In this work we consider the variational, and approximate, search of the ground state of the Hamiltonian within a family of variational states $|\psi_\theta\rangle$, where each member of the variational family is characterized by a set variational parameters $\{\theta \}$. Some examples of variational states include the physically motivated Slater-Jastrow {\it ansatz}~\cite{ceperley1977monte}, the recently introduced neural quantum states~\cite{Carleo2017Science} or unitary quantum circuits~\cite{Peruzzo2014OG-VQE, Kandala2017VQE, CunhaMotta2022VQEbenchmark}. The variational principle establishes that the ground-state energy $E_0$ is a lower bound for the expectation value of the Hamiltonian over the variational family parametrized by $\{\theta \}$:
\begin{equation}
    E_0 \leq \frac{\langle \psi_\theta | \hat{H} | \psi_\theta\rangle}{\langle \psi_\theta | \psi_\theta\rangle}.
\end{equation}
The optimization of this bound yields the optimal set of parameters $\{\bar{\theta}\}$ for which $|\psi_{\bar{\theta}} \rangle$ is the closest approximation to the ground state. Within a fixed basis-set, $\{\bar{\theta}\}$ is strongly dependent on the choice of single-particle basis used to express the Hamiltonian of Eq.~\ref{Eq01:Hmailtonian}. 

It is common practice in quantum chemistry to use basis sets consisting on non-orthogonal and hydrogen-like atomic orbitals as the reference basis. Typically, variational post Hartree-Fock methods such as VMC or VQE use the orthogonal Hartree-Fock orbitals, also known as MOs, as the reference basis to write the Hamiltonian of Eq.~\ref{Eq01:Hmailtonian}~\cite{xia2018_OG-NQS-QChem,Choo2020Chemistry, CunhaMotta2022VQEbenchmark}. Furthermore, the integrals $h_{pq}$ and $h_{pqrs}$ that can be obtained from standard quantum chemistry software packages~\cite{Qiming2020Pyscf} are oftentimes expressed in the MO basis.

While MOs provide an intuitive framework to motivate post Hartree-Fock improvements and study low-laying excitations, this choice may not be favorable for any member of the variational family $|\psi_\theta \rangle$ to accurately represent the ground-state. Therefore, it is desirable to find the optimal single-particle basis that allows for the optimal approximation of the target ground state, by a member of the variational family.

\subsection{Single-particle orbital rotations}
Single-particle orbital rotations transform the Hamiltonian in Eq.~\ref{Eq01:Hmailtonian} via the similarity transformation
\begin{equation}\label{Eq03:similaryTransform}
    \hat{\tilde{H}} = U^\dagger(\kappa) \hat{H} U(\kappa)
\end{equation}
with 
\begin{equation}\label{Eq04:rotation parametrization}
    U(\kappa) = \exp\left( \sum_{\substack{pq \\ \sigma}} \kappa_{pq} \hat{c}^\dagger_{p\sigma} \hat{c}_{q\sigma}\right).
\end{equation}
$\kappa \in \mathbb{R}^{N_{\textrm{orb}} \times N_{\textrm{orb}}}$ parametrizes the above unitary transformation. Unitarity is preserved as long as $\kappa_{pq} = -\kappa_{qp}$. Thouless's theorem~\cite{Thouless1960} establishes that the similarity transformation defined by $U(\kappa)$ turns single- and two-particle operators into  single- and two-particle operators respectively:
\begin{equation}\label{Eq05:CreationRotation}
    \hat{c}^\dagger_{p\sigma} \mapsto \hat{\tilde{c}}^\dagger_{p \sigma} =  \sum_{i} \Phi_{i p} \hat{c}^\dagger_{i\sigma},
\end{equation}
where $\Phi\in \mathbb{R}^{N_{\textrm{orb}} \times N_{\textrm{orb}}}$ is the matrix exponential of $\kappa$: $\Phi = \exp (\kappa)$. The Hamiltonian in the rotated basis can therefore be written in terms of the reference creation and annihilation operators as
\begin{equation} \label{Eq06:RotatedHmailtonian}
    \begin{split}
    \hat{\tilde{H}} = &  \sum_{\substack{pq \\ \sigma}} \tilde{h}_{pq} \hat{c}^\dagger_{p\sigma} \hat{c}_{q\sigma} + \\
    & \sum_{\substack{pqrs \\ \sigma}}  \frac{\tilde{h}_{pqrs}}{2}\left( \hat{c}^\dagger_{p\sigma} \hat{c}^\dagger_{q\sigma} \hat{c}_{s\sigma}  \hat{c}_{r\sigma} +
     \hat{c}^\dagger_{p\sigma} \hat{c}^\dagger_{q -\sigma} \hat{c}_{s -\sigma}  \hat{c}_{r \sigma}\right)
    \end{split}
\end{equation}
where the integrals have been transformed via the tensor transformations
\begin{equation}\label{Eq07:RotatedIntegrals}
    \begin{split}
        \tilde{h}_{pq}  & = h_{ij} \Phi_{i p} \Phi_{j q} , \\
        \tilde{h}_{pqrs}  & = h_{ijkl} \Phi_{i p} \Phi_{j q} \Phi_{k r} \Phi_{l s},
    \end{split}
\end{equation}
where Einstein's summation convention is used.
This type of similarity transformation is very practical as it allows to rotate to the basis that diagonalizes the non-interacting Hamiltonian. Furthermore, in periodic systems allows to switch between plane-wave and localized bases and for quantum chemistry applications allows to define active spaces with the relevant degrees of freedom that intervene on chemical processes~\cite{roos1980complete}. 


\subsection{Variational optimization of the single-particle basis together with the variational state\label{Sec2b}}
A variational procedure is introduced for the search of the single-particle basis that allows for the optimal description of the ground state, given the \textit{bare} variational trial state $|\psi_\theta \rangle$. 

Consider the variational state \textit{dressed} by the single-particle orbital rotations 
\begin{equation}\label{Eq08:variationalRotated}
    |\psi_{\{ \kappa, \theta\}}\rangle = U(\kappa) |\psi_\theta \rangle.
\end{equation}
The loss function to be optimized is defined by the Rayleigh quotient
\begin{equation}\label{Eq09:rotatedCostFunction}
\mathcal{L} (\kappa, \theta) = \frac{\langle \psi_\theta | U^\dagger(\kappa) \hat{H} U(\kappa) | \psi_\theta\rangle}{\langle \psi_\theta | \psi_\theta\rangle} = \frac{\langle \psi_\theta |  \hat{\tilde{H}} (\kappa) | \psi_\theta\rangle}{\langle \psi_\theta | \psi_\theta\rangle}.
\end{equation}

Gradient descent and its variants are used to minimize $\mathcal{L} (\kappa, \theta)$. Gradients with respect to $\{ \theta \}$ are given by the gradients of the expectation value of Hamiltonian in the rotated basis with respect to the bare variational {\it ansatz}:
\begin{equation}\label{Eq10:HamiltonianGradients}
    F_\theta = \frac{\partial }{\partial \theta} \left( \frac{\langle \psi_\theta |  \hat{\tilde{H}} (\kappa) | \psi_\theta\rangle}{\langle \psi_\theta | \psi_\theta\rangle} \right).
\end{equation}
Gradients with respect to the orbital rotations can be computed by the contraction of the bare one- and two-body reduced density matrices (1- and 2-RDMs) with the gradients of the one- and -two body integrals with respect to $\kappa_{pq}$:
\begin{equation}
\begin{split}
    F_{\kappa} & = \frac{\partial}{\partial  \kappa} \left(  \frac{\langle \psi_\theta |  \hat{\tilde{H}} (\kappa) | \psi_\theta\rangle}{\langle \psi_\theta | \psi_\theta\rangle} \right) \\
    & = \sum_{\substack{pq \\ \sigma}} \tilde{h}_{pq}' \Gamma_{pq}^{\sigma} + 
     \sum_{\substack{pqrs \\ \sigma}}  \frac{\tilde{h}_{pqsr}'}{2}\left( \Gamma_{pqrs}^{\sigma, \sigma} +\Gamma_{pqsr}^{\sigma, -\sigma}\right).
    \end{split}
\end{equation}
In the above expression the gradients of the integrals in the rotated basis are given by
\begin{equation}
    \begin{split}
        \tilde{h}_{pq}'  & = h_{ij}\frac{\partial}{\partial \kappa} \left( \Phi_{i p} \Phi_{j q} \right)\\
        \tilde{h}_{pqrs}'  & = h_{ijkl} \frac{\partial}{\partial \kappa} \left( \Phi_{i p} \Phi_{j q} \Phi_{k r} \Phi_{l s}\right),
    \end{split}
\end{equation}
and the bare 1-RDMs and 2-RDMs are defined as
\begin{equation}
    \begin{split}
        \Gamma_{pq}^\sigma & = \frac{\langle \psi_\theta |\hat{c}^\dagger_{p \sigma} \hat{c}_{q \sigma}| \psi_\theta \rangle}{\langle \psi_\theta | \psi_\theta \rangle} \\
        \Gamma_{pqsr}^{\sigma, \sigma '} & = \frac{\langle \psi_\theta |\hat{c}^\dagger_{p \sigma} \hat{c}^\dagger_{q \sigma '} \hat{c}_{s \sigma '} \hat{c}_{r \sigma} | \psi_\theta \rangle}{\langle \psi_\theta | \psi_\theta \rangle}. 
    \end{split}
\end{equation}
The gradients of the integrals with respect to the rotation parameters are computed using the automatic differentiation (AD) tools of the software package Jax~\cite{jax2018}.

Updates of the variational parameters are carried out according to the stochastic reconfiguration method~\cite{Sorella1998StochasticReconfiguration, sorella2007StochasticReconfiguration}, an extension of the classical natural gradient optimization technique~\cite{amari1998natural} for the variational search of quantum states. In the case of $|\psi_\theta \rangle$ being parametrized by a quantum circuit, the quantum natural gradient method is used~\cite{Stokes2020quantumnatural}. Natural gradient approaches depart from the vanilla gradient descent assumption that the parameter space defined by $\{\kappa, \theta \}$ is Euclidean. A more natural choice is to use the Fubini-Study metric to define distances in parameter space. In this case the infinitesimal squared line element is specified by the Fubini-Study metric tensor $g_{ij}(\{\kappa, \theta \}) = 2 \textrm{Re} [ G_{ij} (\{\kappa, \theta \})]$, with
\begin{equation}
\begin{split}
    G_{ij}( & \{\kappa, \theta \})  = \Bigg\langle \frac{\partial \psi_{\{\kappa, \theta\}}}{\partial \{\kappa, \theta\}_i} \Bigg| \frac{\partial \psi_{\{\kappa, \theta\}}}{\partial \{\kappa, \theta\}_j} \Biggl \rangle - \\
    &\Bigg\langle \frac{\partial \psi_{\{\kappa, \theta\}}}{\partial \{\kappa, \theta\}_i} \Bigg|\psi_{\{\kappa, \theta\}} \Biggl \rangle \Bigg\langle \psi_{\{\kappa, \theta\}} \Bigg| \frac{\partial \psi_{\{\kappa, \theta\}}}{\partial \{\kappa, \theta\}_j} \Biggl \rangle.
\end{split}
\end{equation}
While the above definitions assume $|\psi_{\{\kappa, \theta \}} \rangle$ is normalized, they can be extended to non-normalized states~\cite{sorella2007StochasticReconfiguration}. Given the metric tensor $g_{ij}(\{\kappa, \theta \})$, the update of variational parameters $\delta \{\kappa, \theta\}_i$ on each gradient-descent step is given by the solution of the system of equations
\begin{equation}
    \sum_j g_{ij}(\{\kappa, \theta \}) \delta \{\kappa, \theta\}_j = -\eta F_{\{\kappa, \theta \}_i},
\end{equation}
with $\eta$ being the step size. Computing the entries of the metric tensor corresponding to the variational parameters $\kappa$ of the basis rotations is a challenging task. Therefore, we will employ a block diagonal approximation of the metric tensor,
\begin{equation}
    G(\{\kappa, \theta\}) \simeq 
    \begin{blockarray}{cccc}
    \; & \; & \kappa &  \theta   \\
    \begin{block}{c c [cc]}
      \kappa & \;\;& \boxed{\identity} & \;  \\
      \theta & \;\;& \; &  
      \boxed{G(\theta)}  \\
    \end{block}
\end{blockarray},
\end{equation}
where $\identity$ labels the identity matrix. Improvements on the block-diagonal structure of the quantum geometric tensor will be the topic of future investigations. 

The original CAS-SCF formulation was based on the expansion of the energy functional in Eq.~\ref{Eq09:rotatedCostFunction} to either first or second order in the change of $\kappa$, leading into a coupled system of linear equations in the change or orbital-rotation parameters and CI expansion coefficients~\cite{roos1980complete}. This accounts for a crude approximation of the matrix exponential in Eq.~\ref{Eq04:rotation parametrization}. Later improvements~\cite{werner1985second} proposed to expand the cost function in Eq.~\ref{Eq09:rotatedCostFunction} to second order in $\identity-U(\kappa)$, defining the updated orbital rotations by $U(\kappa) \cdot U(\delta \kappa)$. This produces a set of non-linear equations whose solution gives the optimal $\delta \kappa$ to second order. After each $\delta \kappa$ update, the CI coefficients are updated, leading to the alternating optimization of the orbital rotations and CI coefficients. The joint optimization procedure that we propose differs from the technique in Ref.~\cite{werner1985second} in that the wave function and orbital parametrizations are optimized simultaneously. Furthermore, our approach leverages modern AD techniques and Pad\'{e} approximations of the matrix exponential~\cite{Mohy2010Pade-Exponential} to circumvent the need to solve a non-linear system of equations, at the expense of the orbital-rotation updates being correct only to first order in $\delta \kappa$. 

\subsection{Consequences in Variational Monte Carlo}
Variational Monte Carlo requires the definition of the variational state amplitudes $\psi_\theta (n)$ on a given single-particle basis
\begin{equation}\label{Eq17:VMC ansatz}
    |\psi_\theta \rangle = \sum_n \psi_\theta(n) |n\rangle
\end{equation}
with $|n\rangle = \prod_i \left( \hat{c}^\dagger_i \right)^{n_i} |0\rangle$, where $n_i$ is the $i^{\textrm{th}}$ element of the bit-string  $n\in \{0, 1\}^{2N_{\textrm{orb}}}$ whose $i^{\textrm{th}}$ entry specifies the occupancy of each fermionic mode $\hat{c}^\dagger_i$. From now on, the index $i$ summarizes the spin and orbital indices. Given that the anti-symmetry requirements of the fermionic problem has been translated into the anti-commuting algebra of fermionic operators by the Jordan-Wigner mapping, the wave function amplitudes are symmetric functions of the fermionic mode occupancies $n$. Therefore, $\psi_\theta$ can be understood as a function that maps bit-strings of electronic configurations into complex numbers:
\begin{equation}
    \psi_\theta : \{0, 1\}^{2N_{\textrm{orb}}} \mapsto \mathbb{C}.
\end{equation}

Within the VMC framework, the expectation value of an arbitrary operator $\hat{O}$ is estimated as the empirical average of $N_\textrm{samp}$ samples of the so-called local operator $O_\textrm{loc}(n)$ evaluated on sampled basis elements $|n\rangle$:
\begin{equation}
    \frac{\langle \psi_\theta | \hat{O} | \psi_\theta \rangle}{\langle \psi_\theta  | \psi_\theta \rangle} \approx \frac{1}{N_\textrm{samp}} \sum_{n\sim |\psi_\theta (n)|^2}  O_\textrm{loc}(n),
\end{equation}
where $O_\textrm{loc}(n) = \langle n | \hat{O} | \psi_\theta \rangle / \psi_\theta (n)$, and the bit-strings $n$ are sampled using  Markov-Chain Monte Carlo, according to the probability distribution defined by the modulus squared of the wave function amplitudes. With this technique, we obtain unbiased estimates of the expectation value of the Hamiltonian an its gradients with respect to the variational parameters~\cite{Sorella2017QMCBible}. It also allows to obtain unbiased estimates of the metric tensor $G(\theta)$~\cite{sorella2007StochasticReconfiguration}. 

According to Thouless's theorem, the consequence of dressing the variational state in Eq.~\ref{Eq17:VMC ansatz} with single-particle orbital rotations (as defined in Eq.~\ref{Eq08:variationalRotated}) is the continuous and parametrized rotation of the basis used to specify the wave function amplitudes:
\begin{equation}\label{Eq19:VMCrotated}
    |\psi_{\{ \kappa, \theta\}} \rangle = \sum_n \psi_\theta(n) U(\kappa)|n\rangle = \sum_n \psi_\theta(n) |\tilde{n} (\kappa)\rangle,
\end{equation}
where $|\tilde{n} (\kappa)\rangle = \prod_i \left( \hat{\tilde{c}}^\dagger_i (\kappa) \right)^{n_i} |0\rangle$, and the rotated creation operators $\hat{\tilde{c}}^\dagger_i (\kappa)$ are defined as in Eq.~\ref{Eq05:CreationRotation}. This means that the choice of the single-particle basis that defines the trial state is part of the variational freedom of the problem. Furthermore, as described in section~\ref{Sec2b}, the rotation parameters $\kappa$ and the parameters that define the wave function amplitudes $\theta$ can be treated in equal footing and optimized using gradient-based methods.

\section{Numerical experiments}
In this section we present numerical evidence showing that finding the optimal basis for a given variational {\it ansatz} is advantageous on a variety of many-body problems including \emph{ab initio} Hamiltonians, as well as fermionic lattice models. Two classes of variational techniques are considered, VMC with neural quantum states and VQE with hardware-efficient circuits.

\subsection{Variational Monte-Carlo with Neural Quantum States}

NQS refers to the use of neural networks to represent the amplitudes of the variational state $\psi_\theta(n)$ given the bit-string $n$ of of fermionic mode occupancies. NQS for the representation of fermionic states can be separated into three groups: non-determinant based~\cite{xia2018_OG-NQS-QChem, Choo2020Chemistry, Yoshioka2021solids, bennewitz2021VMC+VQE, Barrett2022RNNQuantumChem, Inui2021determinantfree, sureshbabu2021_NQSQC, zhao_2022autoregressiveNQS, yang2020_NQSQChem, Booth2023GPS}, determinant based~\cite{Nomura2017slaterRBM, Riggeri2018IterativeBackflow, Luo2019backflow, Stokes2020, Pfau2020Ferminet, Spencer2020better, Hermann2020Paulinet, RobledoMoreno2022Hidden-Fermions, Glehn2022Psiformer, wu2023varbench}, and hybrid classical-quantum \textit{ans\"atze}~\cite{barison2023embedding, huembeli2022entanglementForgingNQS}. 

The non-determinant based NQS amplitudes $\psi_\theta(n)$ are the output of a feedforward neural network that takes as an input the bit-string of occupancies $n$. Two different architectures are considered here. The first is a fully connected network with a single hidden layer. From now on this architecture is referred to as feed-forward network (FFN). The activation function of the hidden layer is the rectified linear unit and the activation function of the output layer is the hyperbolic tangent ($\tanh$). The number of hidden units in the hidden layer is determined by the product $\alpha \cdot 2N_{\textrm{orb}}$, where $\alpha$ is a hyper-parameter that controls the expressive power of the neural network. The second architecture is the Restricted Boltzmann Machine (RBM). The RBM has  been extensively used to parametrize wave functions of both spin~\cite{Carleo2017Science} and fermionic~\cite{xia2018_OG-NQS-QChem, sureshbabu2021_NQSQC, Choo2020Chemistry, Yoshioka2021solids, Nomura2017slaterRBM} systems. The expressive power of the RBM is also determined by the number of hidden units, which similarly to the FFN, can be characterized by the hyper-parameter $\alpha$. We set $\alpha = 16$ in all cases. The variational weights of the FFN are taken to be real while the variational parameters of the RBM are taken to be complex. This is because the RBM of real parameters only produces positive real outputs.

The determinant-based variational states are composed of the product of a $N\times N$ determinant (where $N$ is the total number of electrons) and optionally a parametrized symmetric function of orbital occupancies (Jastrow factor $J:n \mapsto \mathbb{R}$)
\begin{equation}
    \psi_\theta(n) = \det \big[ \bar{n} \cdot  \Xi \big] \; J(n),
\end{equation}
where $\bar{n} \in \{0, 1\}^{N \times 2N_{\textrm{orb}}}$ is the matrix with $\bar{n}_{ij} = 1$ for the $i^{\textrm{th}}$ occupied orbital in the $j^\textrm{th}$ component of $n$ and zero otherwise. $\Xi \in \mathbb{R}^{2N_{\textrm{orb}} \times N}$ is what is known as the matrix of orbitals. The product $\bar{n} \cdot  \Xi$ produces a $N\times N$ matrix obtained from $\Xi$ by including row $j$ of $\Xi$ if $n_j = 1$. If $\Xi$ is a matrix of independent variational weights, then $\det \big[ \bar{n} \cdot  \Xi \big]$ is a Slater determinant. If the matrix elements of $\Xi$ are parametrized functions of $n$, then $\det \big[ \bar{n} \cdot  \Xi(n) \big]$ is a determinant with backflow correlations. Determinant-based {\it ans\"{a}tze} are commonly used in a first-quantized description of the problem due to the anti-symmetry introduced by the permutation of rows of the determinant~\cite{Luo2019backflow, Pfau2020Ferminet, Spencer2020better, Glehn2022Psiformer, Hermann2020Paulinet}. However, as described above, determinant-based {\it ans\"{a}tze} can be adapted to work in a second-quantized description of the problem, obtaining a $\psi_\theta$ that is a symmetric function of the occupancy of the fermionic modes.

Two different determinant-based variational states have been considered. The first one is a Slater determinant (no Backflow dependence in $\Xi$) multiplied by a Jastrow factor parametrized by a fully connected neural network with a single hidden layer with $\tanh$ activation functions. The width of the hidden layer is $\alpha = 16$. This variational state is referred to as Slater NN Jastrow from now on. The second determinant-based state consists on a Backflow determinant, with no Jastrow factor, where each row of $\Xi(n)$ is parametrized by a unique single hidden-layer fully-connected neural network with $\tanh$ activation functions and width $\alpha = 16$. This variational state is referred to as NN Backflow from now on. Given that a simple Slater determinant can be interpreted as the action of a basis rotation on top of an uncorrelated reference state, it might appear that variational approaches employing determinant-based {\it ans\"{a}tze} are invariant under the reparametrization of the Hamiltonian by the orbital rotations.  However, when the determinant includes electronic correlations via a Jastrow factor or Backflow corrections this is not the case anymore.

Four systems are studied with the different variational states described above. The first one is the $\textrm{H}_4$ molecule where hydrogen atoms are placed in the vertices of a rectangle with fixed diagonal length but variable proportions amongst the short and long sides. The second one is the six-site one-dimensional Hubbard model with periodic boundary conditions at half filling. The third one is the dissociation of the $\textrm{BH}$ molecule. All of the previously mentioned molecules are described in the minimal $\textrm{STO-6G}$ basis set. 
The fourth system is the ethylene molecule, described in the correlation-consistent cc-pVDZ basis set. In this case we consider a highly compact class of NQS variational wave functions based on an active space formalism. The search for the ground state of the previously mentioned molecules is performed at charge neutrality, where the total number of electrons $N$ equals the number of protons, and in the subspace of total zero magnetization. All VMC calculations are implemented using the NetKet software package~\cite{netket32022}. 

\begin{figure*}
    \includegraphics[width = \textwidth]{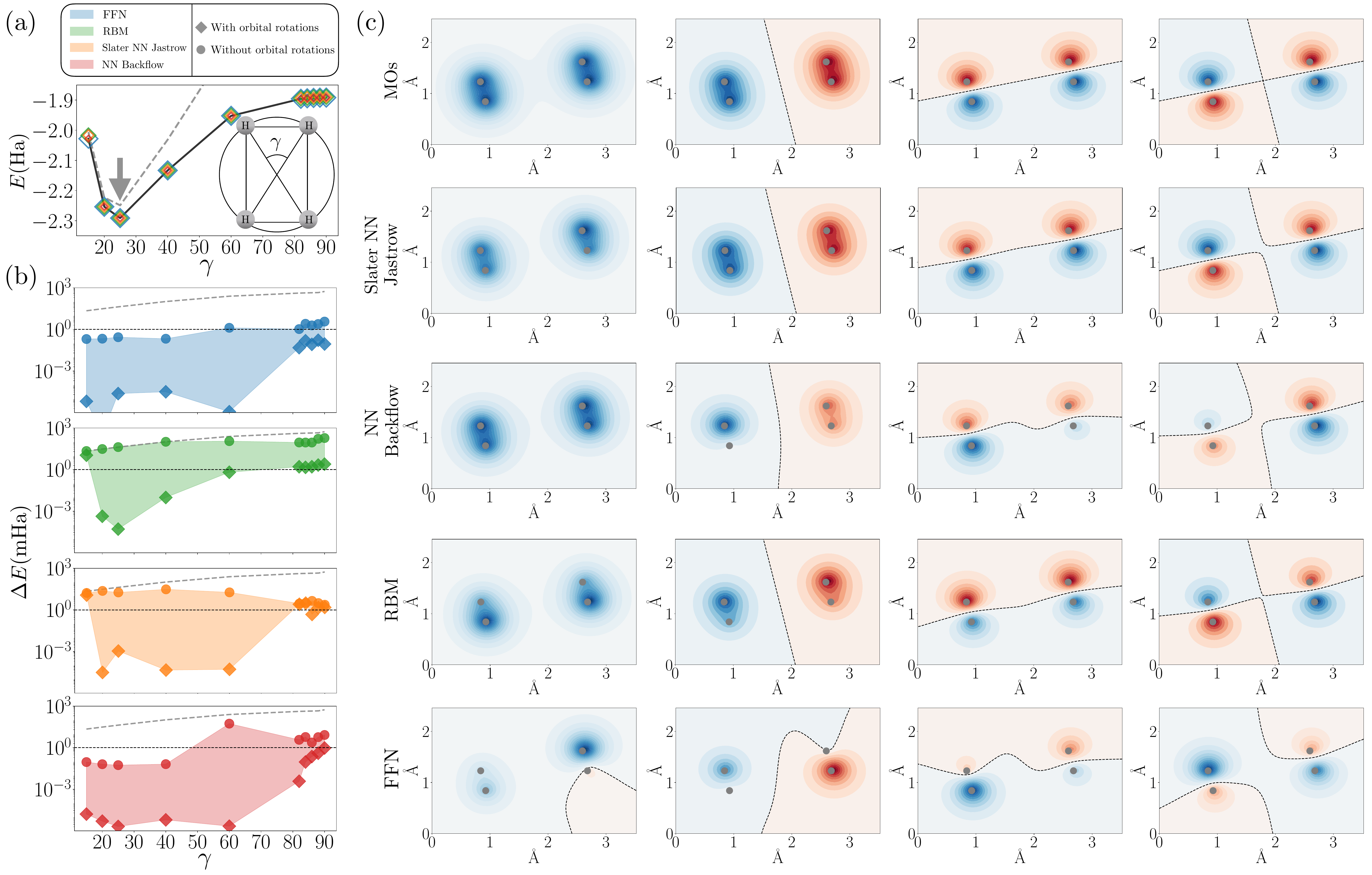}
    \caption{\label{fig_01:H4} Variational Monte Carlo results using NQS states with single-particle orbital optimizations in the $\textrm{H}_4$ molecule described in the minimal STO-6G minimal basis set. All four hydrogen atoms form a rectangle whose diagonal has a fixed length of $3.475953(1) \; \textrm{\AA}$.  \textbf{(a)} Ground-state energy as a function of the internal angle $\gamma$ parametrizing the geometry of the molecule. The dashed grey and solid black curves indicate the Hartree-Fock and exact energies respectively. The diamonds correspond to the energies obtained from VMC using orbital optimizations during the energy minimization. Different colors correspond to different NQS {\it ans\"{a}tze}. \textbf{(b)} Difference between the exact and variational ground state energies for different variational states. Dots correspond to the converged energies in the fixed MO basis while diamonds are the results when the optimization includes orbital rotations. The shaded region highlights the improvement in the errors when orbitals are optimized. \textbf{(c)} Each row shows the four converged orbitals for $\gamma = 25 ^\circ$, in the plane where the atoms live,  after their optimization with the corresponding variational {\it ansatz}. The first row shows the molecular orbitals for reference. The blue scale shows positive values while the red scale shows negative values. The black dashed line shows the nodal surfaces. The grey dots indicate the position of the Hydrogen atoms.}
\end{figure*}

\begin{figure}
    \includegraphics[width = 1 \columnwidth]{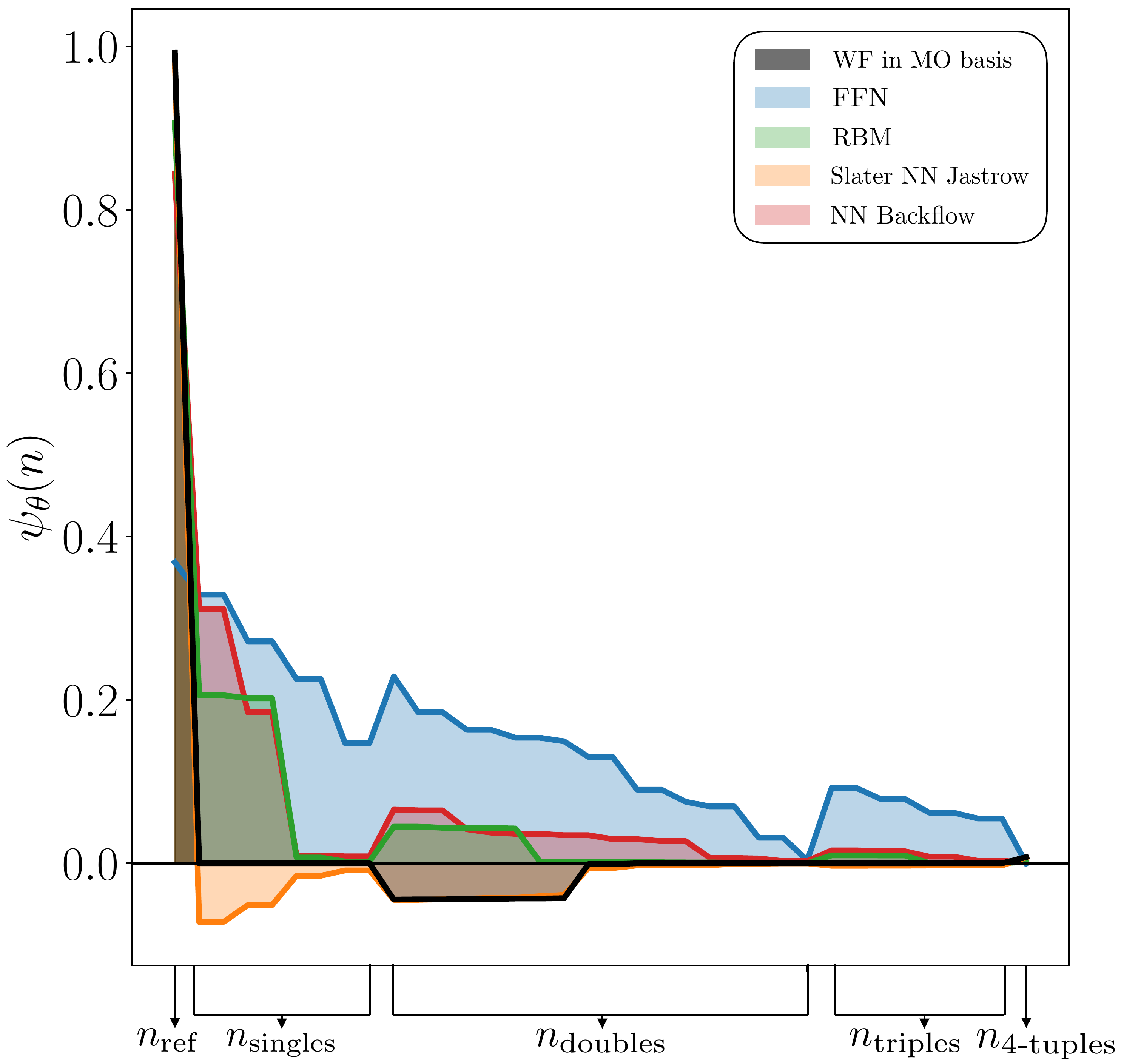}
    \caption{\label{fig_01_p:H4_amplitudes} Ground-state wave function amplitudes for the $\textrm{H}_4$ molecule in the minimal STO-6G minimal basis set at $\gamma = 25^\circ$. Different curves labelled by their color show the wave function amplitudes in the optimal single-particle basis (as shown in Fig.~\ref{fig_01:H4} (c)) for the FFN, RBM, Slater NN Jastrow and NN Backflow variational states. The ground-state wave function amplitudes in the MO basis are shown for reference. $n_\textrm{ref}$ is the bit-string with the largest wave function amplitude. $n_\textrm{singles}$, $n_\textrm{doubles}$, $n_\textrm{triples}$ and $n_\textrm{4-tuples}$ refer to the occupancy bit-strings whose half-Hamming distance to $n_\textrm{ref}$ is 1, 2, 3 and 4 respectively.}
\end{figure}

\paragraph*{\textbf{$\textrm{H}_4$ molecule.--}} We consider the arrangement of four Hydrogen atoms in the vertices of a rectangle of fixed diagonal length $3.475953(1) \; \textrm{\AA}$. The geometry of the molecule is controlled by the angle $\gamma$ between adjacent atoms as shown in the inset of Fig.~\ref{fig_01:H4} (a). This defines a small toy model of strongly correlated electrons, that can be solved exactly. The energy is a smooth function of $\gamma$ with a maximum for $\gamma = 90^\circ$. Due to a level crossing between two Hartree-Fock states at $\gamma = 90^\circ$, approximate methods like Coupled Cluster with singles and doubles, and triples treated perturbatively, (CCSD(T)) fail to describe a physical energy landscape~\cite{VanVoorisH4}. At $\gamma = 90^\circ$, CCSD(T) predicts a local minimum in the energy with a nonphysical cusp.

Panel (a) of Fig.~\ref{fig_01:H4} shows the comparison of the exact ground state energy with the variational energy as a function of the diagonal angle for the FFN, RBM, Slater NN Jastrow and NN Backflow trial states, including orbital rotations as part of the optimization. All states show good agreement in their variational energies with the exact energy for all angles considered. In particular, close to $90^\circ$ the energy is a smooth function of the angle. Panel (b) in Fig.~\ref{fig_01:H4} shows the improvement in the energy error $\Delta E$ as a function of $\gamma$ when basis rotations are included as part of the variational procedure. In all cases the optimization of the bare and dressed variational {\it ans\"{a}tze} are initialized with the same set $\{ \theta \}$ of variational parameters. In all cases the energies are improved for all $\gamma$ values if orbital optimizations are taken into account. 

Panel (c) in Fig.~\ref{fig_01:H4} shows the converged rotated orbitals for the dressed FFN, RBM, Slater NN Jastrow and NN Backflow as well as the reference MOs for a configuration of atoms close to the equilibrium geometry ($\gamma = 25^\circ$). We observe that the determinant based states experience a lesser alteration of the orbitals that define the basis for the Hilbert space, while the converged orbitals for the non-determinant based {\it ans\"{a}tze} show greater alterations as compared to the reference MOs. Furthermore, non-determinant based {\it ans\"{a}tze} tend to favor single-particle bases whose orbitals are localized around individual Hydrogen atoms. This observation is explained by the fact that the determinant-based {\it ans\"{a}tze} are capable of representing more sparse, in bit-string space, wave function structures, while non-determinant based {\it ans\"{a}tze} seem to be better suited to approximate smooth functions in bit-string space. 

Fig.~\ref{fig_01_p:H4_amplitudes} shows the ground-state wave function amplitudes for the $\textrm{H}_4$ molecule at $\gamma = 25^\circ$ for different choices of single-particle bases. In the MO basis, the Hartree-Fock ground state is predominant over the rest of electronic configurations. In the single-particle bases where the expressive power of the Staler NN Jastrow and NN Backflow {\it ans\"{a}tze} is maximized (see Fig.~\ref{fig_01:H4} (c)), the ground-state amplitudes are sparse and mostly dominated by a single electronic configuration, similar to the MO basis. In the single-particle bases preferred by the RBM and FFN {\it ans\"{a}tze} (see Fig.~\ref{fig_01:H4} (c)), the wave function amplitudes have a less sparse structure in the electronic configuration apace. In particular, the single-particle basis corresponding to the FFN leads to ground-state wave function amplitudes spreading across the whole space of bit-strings of electronic configurations. 

Furthermore, Fig.~\ref{fig_01_p:H4_amplitudes} shows that in the particular case of the $\textrm{H}_4$ molecule at $\gamma = 25^\circ$, there exist multiple single-particle bases for which the ground-state amplitudes are positive semidefinite. We observe that this effect appears in conjunction with the alteration of the nodal surfaces of the single-particle orbitals that define the bases, as shown in Fig.~\ref{fig_01:H4} (c). This observation is relevant in the context of VMC, as capturing non-trivial sign structures has proven to be a challenging task~\cite{Choo2019J1J2, Stokes2020Spinless, Luo2019NNBackflow}. The improvement of challenging sign structures via orbital rotations, on a different flavor of quantum Monte Carlo, has also been reported in Ref.~\cite{Levy2021SingMitigatedBasisRots}. Additionally, Fig.~\ref{fig_01_p:H4_amplitudes} shows that the multi-modality characteristics of the probability distribution defined by $|\psi_\theta(n)|^2$ depends on the choice of single-particle basis. For example, the ground-state amplitudes in the MO basis have a distinct bi-modal structure, where the two modes are separated by moves of two fermions in Fock space. In contrast, the ground-state amplitudes in the basis favorable for the FFN {\it ansatz} has a single-mode structure. This is of extreme relevance in the Markov-Chain sampling of the wave function to compute expectation values. Multi-modal probability distributions represent pathological cases as they require an exponential length of the sampling chains to converge to the true underlying distribution.

\begin{figure*}
    \includegraphics[width = \textwidth]{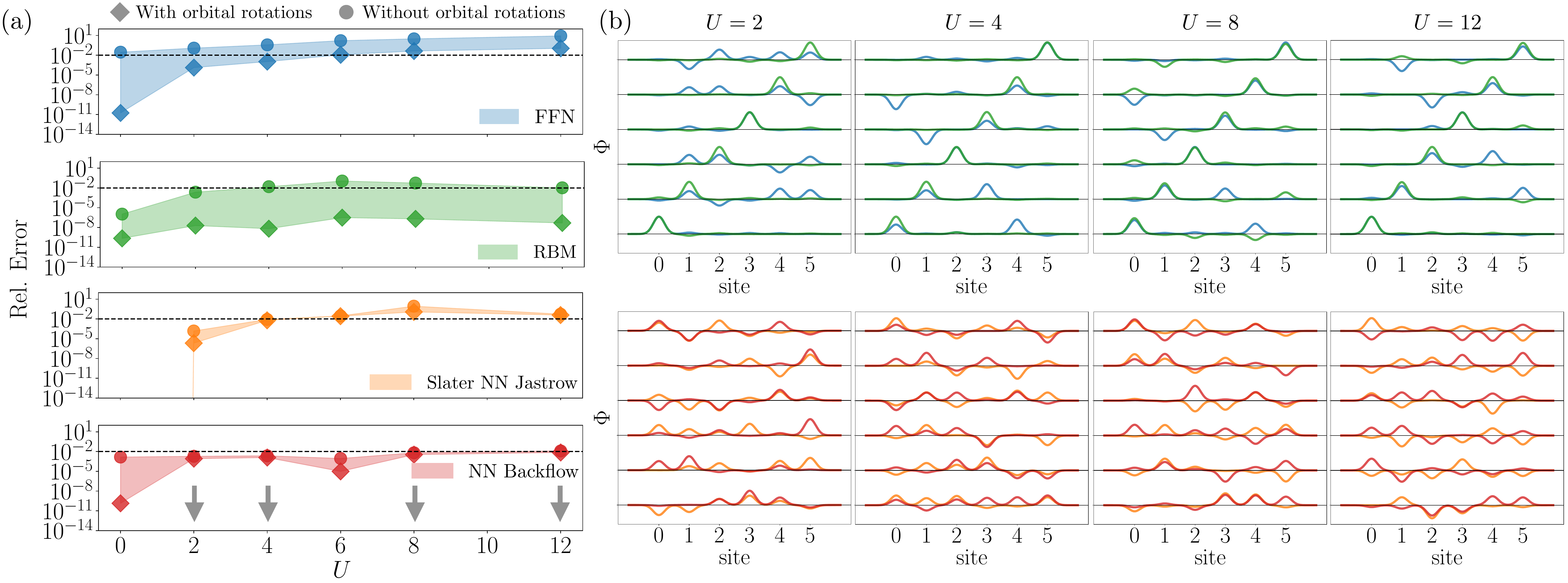}
    \caption{\label{fig_02:Hubbard} Variational Monte Carlo results using NQS {\it ansatz} with single-particle orbital optimizations in the 6-site, one-dimensional Hubbard model with periodic boundary conditions at half filling. \textbf{(a)} Relative error in the ground state energy as a function of the onsite repulsion $U$. Different panels show the error for different classes of NQS. Dots correspond to the converged energies of the fixed localized basis while diamonds are the results when the optimization includes orbital rotations. The shaded region highlights the improvement in the errors when optimizing the single-particle basis. \textbf{(b)} Converged single-particle orbitals after the optimization of the variational state together with the single-particle basis. Different columns correspond to different values of $U$ as indicated. Different colors correspond to the different variational states, matching the color scheme of panel (a). The orbitals have been artificially widened by a Gaussian kernel for visualization purposes. }
\end{figure*}

\paragraph*{\textbf{Hubbard model.--}} The Hubbard model provides a minimal description of locally interacting electrons hopping between adjacent sites of a given lattice. In the so-called localized basis it is given by the Hamiltonian
\begin{equation}
    \hat{H} = -\sum_{\{i, j\}, \sigma} \left( \hat{c}^\dagger_{i\sigma} \hat{c}_{j\sigma} + \hat{c}^\dagger_{i\sigma} \hat{c}_{j\sigma} \right) + U \sum_i \hat{n}_{i\sigma} \hat{n}_{i -\sigma},
\end{equation}
with $\hat{n}_{i\sigma} =  \hat{c}^\dagger_{i\sigma} \hat{c}_{i\sigma}$ being the number operator and $U$ the interaction strength. For large values of $U$, the ground-state is strongly correlated, displaying competing charge and antiferromagnetic orders~\cite{Zheng2017stripes}. The Hubbard model is a particular case of the more general Hamiltonian in Eq.~\ref{Eq01:Hmailtonian}, with very sparse one- and two-body integrals. In particular, the number of non-zero elements of the two-body integrals grows linearly with the number of sites as opposed to the general case where it grows with the fourth power of the number of orbitals. In this section we study the one-dimensional 6-site Hubbard model with periodic boundary conditions at half filling.

While it is an interesting exercise to study the improvement of the variational energy if the basis is rotated from the localized basis to a different single-particle basis, in practice, rotating the Hubbard model to an arbitrary single-particle basis significantly increases the computational cost of the energy optimization. In the localized basis, the number of terms of the Hubbard Hamiltonian is proportional to the number of sites considered. In an arbitrary single-particle basis, the number of terms in the Hamiltonian scales with the fourth power of the number of sites.

Figure~\ref{fig_02:Hubbard} (a) shows the comparison of the relative error in the ground-state energy as a function of the interaction strength $U$, between the bare variational states and the orbital-rotation dressed states. In all cases the optimization of the bare and dressed variational {\it ans\"{a}tze} are initialized with the same set $\{ \theta \}$ of variational parameters. In all cases we observe an improvement of the variational energy when orbital rotations are included as part of the optimization. 

Figure~\ref{fig_02:Hubbard} (b) shows the converged single-particle orbitals where the expressive power of the FFN, RBM, Slater NN Jastrow and NN Backflow {\it ans\"{a}tze} is maximized. At small values of $U$, the single-particle basis preferred by the FFN {\it ansatz} shows a delocalization of some of the orbitals, while others remain localized. At larger values of $U$, a single orbital remains localized, while the rest spread over two non-adjacent sites.

The RBM {\it ansatz}, for all $U$ values favors a mostly localized set of single-particle orbitals. This observation is consistent with the fact that the RBM can efficiently represent any matrix product state (MPS)~\cite{Chen2018RBM=TNS}. Given that any state with area-law scaling entanglement can be efficiently represented by an MPS~\cite{Verstraete2004AreaMPS}, and that gapped one-dimensional systems (like the Hubbard model under consideration) have ground sates that satisfy an area law~\cite{Hastings20071DAreaLaw}, we expect that the ground state of this one-dimensional chain (in the localized basis) can be efficiently represented by and MPS and therefore by an RBM. We also remark that while the rotated orbitals are very close to the localized ones, the energy error is reduced when basis rotations are included (see Fig.~\ref{fig_02:Hubbard} (a)). Therefore, we conclude that in this particular case, the improvement in the variational energy is mostly due to a more favorable energy landscape in parameter space when orbital rotations are included in the optimization procedure. This observation is discussed in mode detail in Sec.~\ref{Sec:VQE}. 

In contrast to the non-determinant based {\it ans\"{a}tze}, the expressive power for determinant-based {\it ans\"{a}tze} is observed to be maximized for delocalized single-particle orbitals for all values of $U$.


\begin{figure*}
    \includegraphics[width = \textwidth]{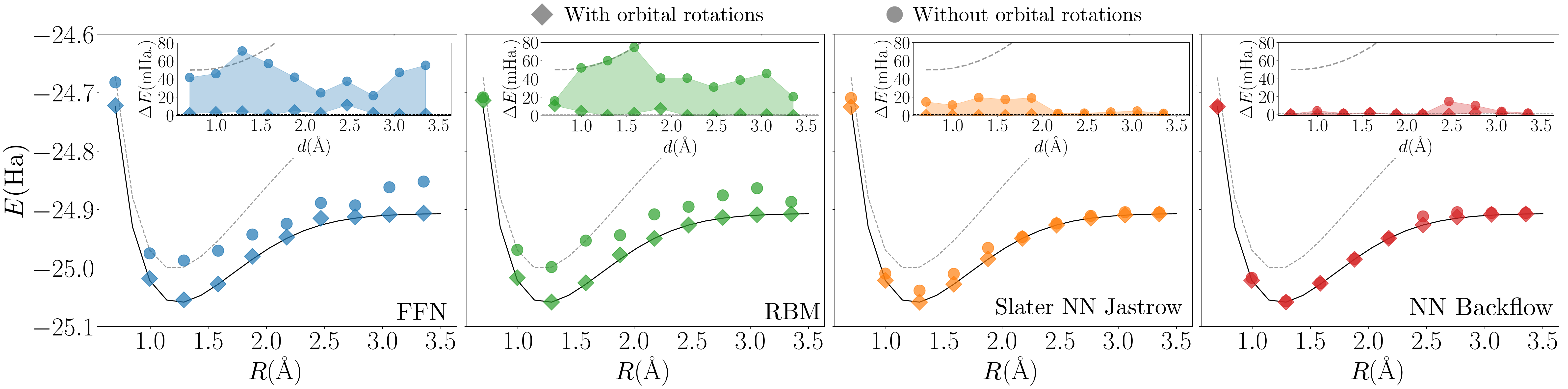}
    \caption{\label{fig_03:BH} Variational Monte Carlo energies and energy differences obtained using NQS with orbital optimizations in the \textrm{BH} molecule  described in the minimal STO-6G minimal basis set. Each panel shows the energy as a function of bond length for different NQS {\it ansatz}. The dashed grey and solid black curves indicate the Hartree-Fock and exact energies respectively. The inset shows the difference between the exact and variational energies. Dots correspond to the converged energies of the fixed MO basis while diamonds are the results when the optimization includes orbital rotations. The shaded region highlights the improvement in the errors when optimizing the single-particle basis.}
\end{figure*}

\paragraph*{\textbf{$\textrm{BH}$ molecule.--}} The dissociation of the covalent bond of the $\textrm{BH}$ molecule provides a good test-bed for variational approaches given the presence of predominantly dynamic (static) correlations at small (medium to large) bond lengths $R$~\cite{CunhaMotta2022VQEbenchmark}. 

Figure~\ref{fig_03:BH} shows the energy and energy error as a function of $R$ in the dissociation of the $\textrm{BH}$ molecule. We observe that the non-determinant based {\it ans\"{a}tze} (FFN and RBM), in the MO basis without orbital rotations, show greater errors in the ground-state energy as compared to the determinant-based {\it ans\"{a}tze} (Slater NN Jastrow and NN Backflow). Like in the $\textrm{H}_4$ molecule case, this caused by the sparse (in bit-string space) representation of the ground-state wave function in the MO basis, where the amplitudes are dominated by the Hartree-Fock ground state. This indicates that determinant-based variational states are better suited to represent sparse functions than non-determinant based trial states. Smooth functions in the bit-string space of electronic configurations are a more favorable scenario for non-determinant based trial states.

As shown in Figure.~\ref{fig_03:BH}, the energy error is decreased when the variational states are dressed with parametrized orbital rotations. The improvement of the energy in the determinant-based sates is less significant than the improvement shown in the non-determinant based states, where energies are improved by tens of $\textrm{mHa}$

\begin{figure*}
    \includegraphics[width = \textwidth]{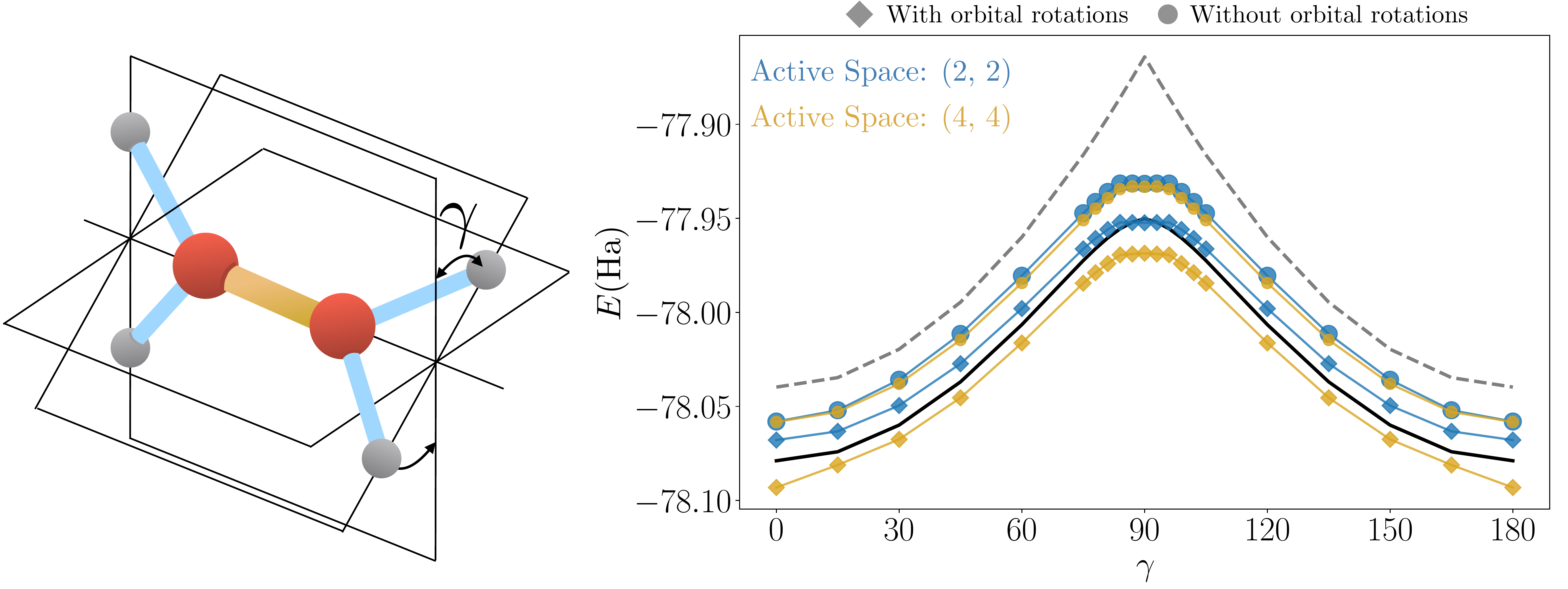}
    \caption{\label{fig_03:Ethy}  Variational Monte Carlo ground state calculation, using NQS, of the ground state properties of the torsion of the ethylene molecule described in the correlation-consistent cc-pVDZ basis set.  \textbf{Left:} Twisting of the ethylene molecule along the $\textrm{C-C}$ bond from its equilibrium configuration ($\gamma = 0^\circ $). Carbon atoms are represented by red spheres while hydrogen atoms are represented by grey spheres. The $\textrm{C-C}$ bond is represented by the orange cylinder and $\textrm{C-H}$ bonds by light blue cylinders. \textbf{Right:} Energy as a function of the torsion angle $\gamma$. The grey dashed line corresponds to the Hartree-Fock energy and the solid black line is the energy of a 12 electron and 12 orbital $(12,12)$ CAS-CI calculation in the fixed MO basis. Active space electrons and orbitals are those closest to the Fermi level. Blue and golden symbols correspond to the variational ground state search in the active space using the NN Backflow \textit{ansatz}, for different active-space sizes as indicated by the different colors. Dots correspond to the converged energies in the fixed MO basis. Diamonds correspond to the converged energies when orbital rotations (amongst all of the orbitals) are included in the optimization. }
\end{figure*}

\paragraph*{\textbf{Torsion of ethylene (active-space calculation).--}} The equilibrium geometry of the ethylene molecule ($\textrm{C}_2\textrm{H}_4$) is shown in the left panel of Fig.~\ref{fig_03:Ethy} when $\gamma = 0$. The $\textrm{C-C}$ bond length is $1.339 \textrm{\AA}$ while the $\textrm{C-H}$ bond lengths are $1.087 \textrm{\AA}$. The angle between the axes defined by the $\textrm{C-C}$ and $\textrm{C-H}$ bonds is $58.7^\circ$. We focus on the energy surface obtained when twisting the molecule along the $\textrm{C-C}$ bond. The molecule is described in the correlation-consistent basis set cc-pVDZ (48 atomic orbitals). correlation-consistent basis sets contain a larger number of basis elements that span the Hilbert space as compared to minimal basis sets, increasing the computational cost of the variational approaches explored so far. 

Consequently, we reduce the computational cost by performing the ground state search in an active space. Consider the subset $\mathcal{A}$ of all allowed electronic configurations  obtained by fixing the occupancy of a set of ``inactive'' spin orbitals to be one and fixing the occupancy of a set of ``virtual'' orbitals to be zero. The remaining electrons (active electrons) are allowed to have any configuration in the remaining set of ``active space'' orbitals. See Appendix~\ref{Appendix: active space energy} for details on the nomenclature of orbitals. The support for the bare variational wave function for an active space calculation is $\mathcal{A}$:
\begin{equation}
    |\psi_\theta \rangle = \sum_{n\in \mathcal{A}} \psi_\theta(n) |n\rangle.
\end{equation}
We see that the variational amplitudes need to be specified over only $|\mathcal{A}|$ electronic configurations. This restricted representation of the wave function reduces the cost of the evaluation of the expectation value of fermionic operators and therefore of the Hamiltonian (see Appendix~\ref{Appendix: active space energy}). If $\psi_\theta(n)$ is given by a lookup table of variational amplitudes, the optimal solution to the problem can be obtained in closed-form from the solution of an eigenvalue problem. This is known as complete active space configuration interaction (CAS-CI). 

In this work we use the NN Backflow \textit{ansatz} to represent the variational state in the active space. In this case, the single-particle basis rotations are applied to all of the orbitals that span the Hilbert space, allowing to mix the active space orbitals with the core and virtual orbitals. Fig.~\ref{fig_03:Ethy} shows the energy as a function of the torsion angle $\gamma$ obtained from different methods. The Hartree-Fock energy shows a cusp at $\gamma = 90^\circ$ degrees due to a degeneracy in the orbital spectrum at the Fermi level. This non-physical behaviour is fixed  in the CAS-CI energy in the fixed MO basis, with an active space with 12 electrons in 12 orbitals $(12, 12)$. The same qualitative behavior is observed when using the NN Backflow \textit{ansatz } with active spaces of two electrons in two orbitals $(2,2)$ and four electrons in four orbitals $(4,4)$ in the fixed MO basis. As expected, since the active spaces are smaller in the NN Backflow calculations compared to the CAS-CI calculation, the energies obtained are higher for all values of $\gamma$. When the bare NN Backflow \textit{ansatz} in the active space is dressed with parametrized orbital rotations, the variational energies are significantly improved as compared to the optimization in the fixed MO basis. It is also important to remark that the $(4,4)$ NN Backflow variational energy obtained when optimizing the single particle basis is lower than the $(12,12)$ CAS-CI energy in the fixed MO basis. This is remarkable since the number of non-zero amplitudes in the $(4, 4)$ active space wave function is $|\mathcal{A}| = 36$ while the number of non-zero amplitudes in the $(12, 12)$ active space wave function amplitudes is $|\mathcal{A}| = 853776$ (noting that we work in the singlet subspace). This observation demonstrates that single-particle basis rotations allow to construct much more compact representation of the interacting-electron wave function.  


\begin{figure}
    \includegraphics[width = 1 \columnwidth]{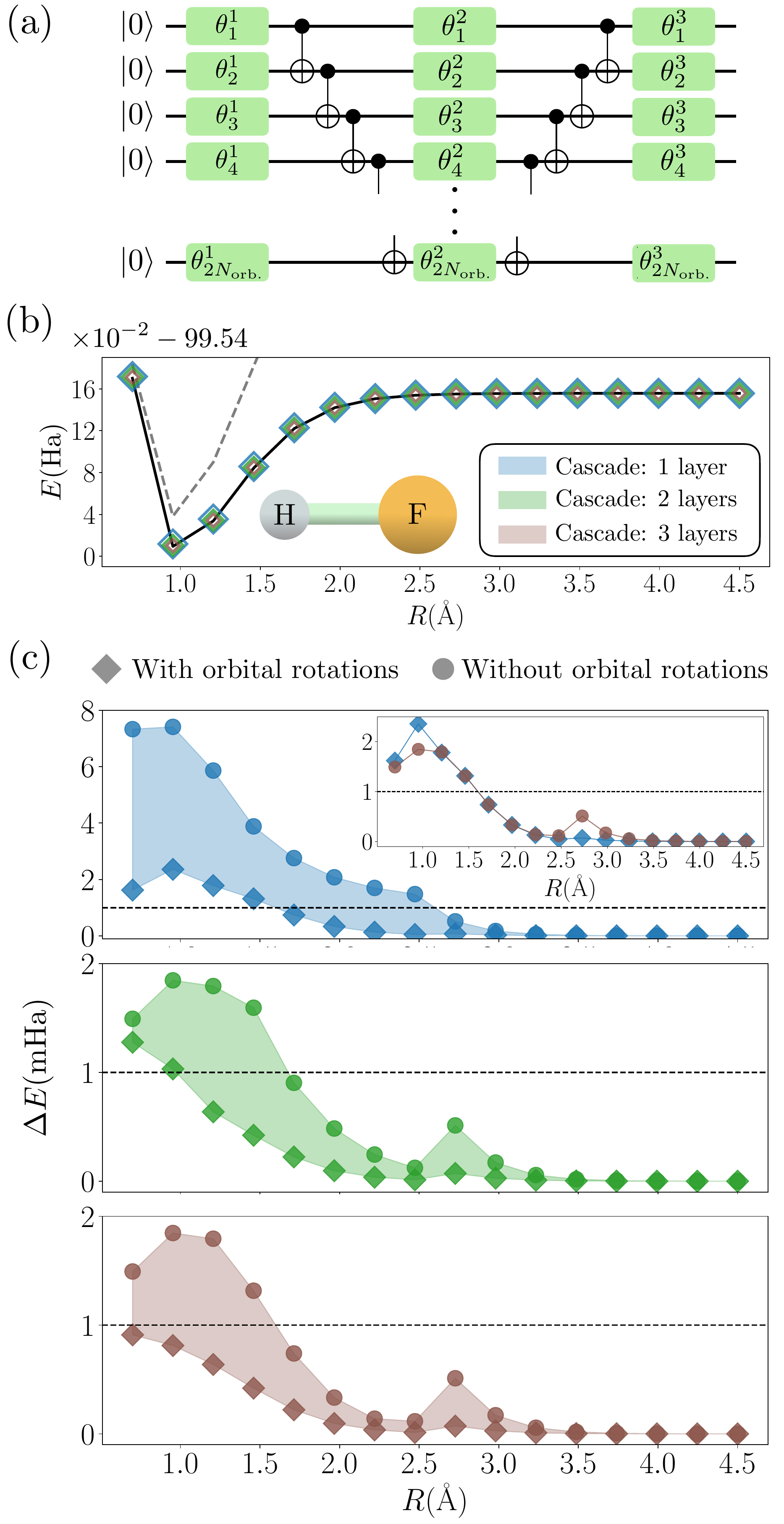}
    \caption{\label{fig_04:HF error}  VQE in the $\textrm{HF}$ molecule in the STO-6G minimal basis set. \textbf{(a)} One layer of the Cascade quantum circuit {\it ansatz}. The green single-qubit gates represent single-qubit rotations along the $\hat{y}$ axis of angle $\theta$. \textbf{(b)} Ground-state energy as a function of $R$ for the Cascade {\it ansatz} with different number of layers dressed with orbital rotations. The black-solid and grey-dashed curves show the exact and Hartree-Fock energies for reference. \textbf{(c)} From top to bottom: ground-state energy error as a function of the bond length $R$ for the one-, two- and three-layer Cascade {\it ansatz}. Circles show the variational error working on the fixed MO basis. Diamonds correspond to the variational error when incorporating orbital rotations. The shaded region highlights the improvement in the energy error. The inset in the top panel compares the error in the ground state energy obtained with three layers of the Cascade {\it ansatz} and no orbital rotations, with a single-layer Cascade {\it ansatz} with orbital rotations.} 
\end{figure}

\begin{figure*}
    \includegraphics[width = \textwidth]{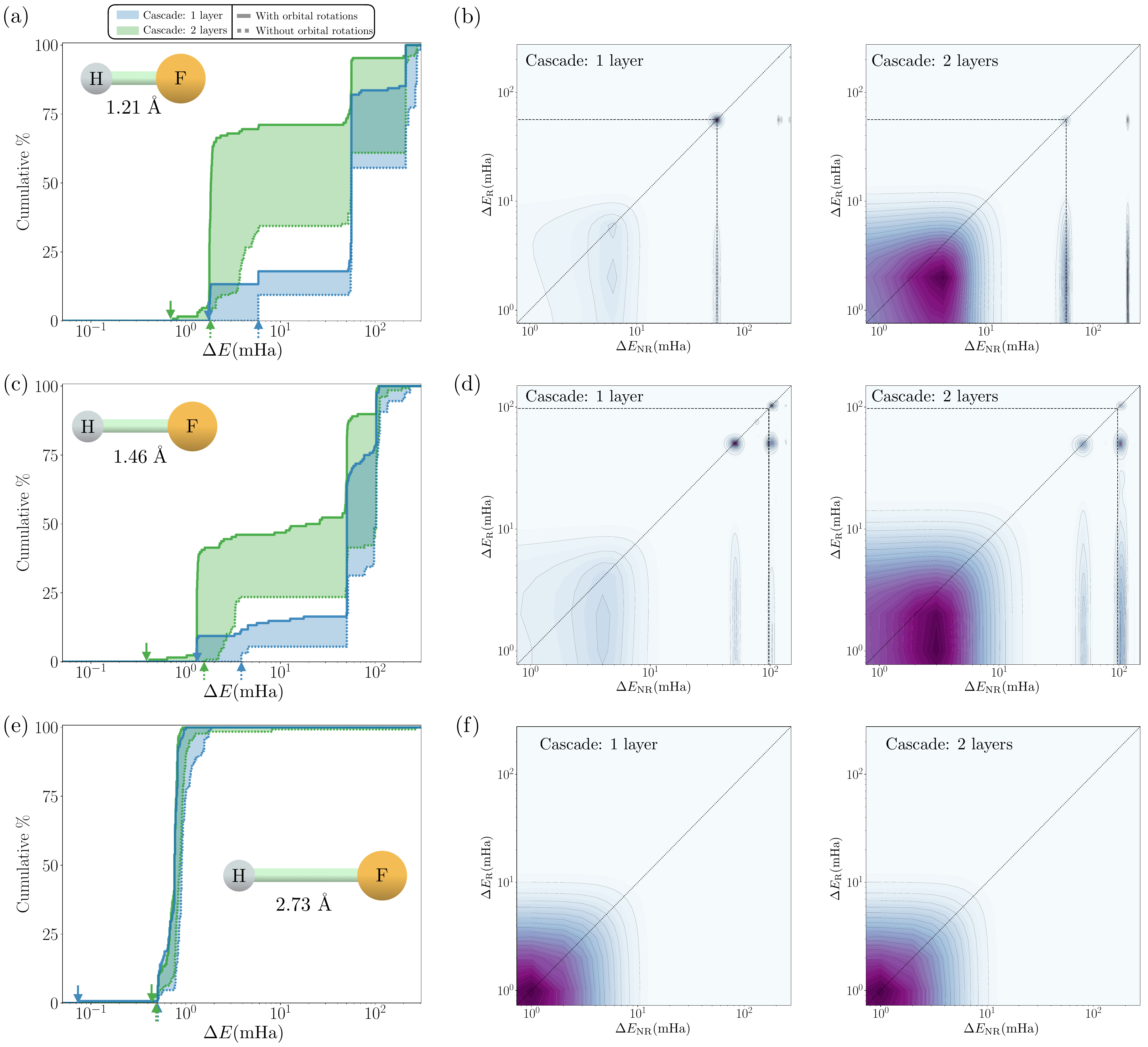}
    \caption{\label{fig_04:HF statistics} VQE optimization empirical statistics for the one- and two-layer cascade {\it ansatz} fapplied to the  $\textrm{HF}$ molecule in the STO-6G minimal basis set at bond lengths $1.21 \; \textrm{\AA}$ (\textbf{(a)}-\textbf{(b)}),  $1.46 \; \textrm{\AA}$ (\textbf{(c)}-\textbf{(d)}) and $2.73 \; \textrm{\AA}$ (\textbf{(e)}-\textbf{(f)}). \textbf{(a)},  \textbf{(c)} and \textbf{(e)} show the cumulative percentage of randomly initialized circuits that converge to a ground-state energy error equal or smaller than $\Delta E$. The blue and green histograms correspond to the one- and two-layer Cascade {\it ans\"{a}tze}. Solid lines show the statistics when incorporating orbital optimizations to the variational procedure while dotted lines correspond to the statistics when the optimization is carried out in the fixed MO basis. The shaded region highlights the difference between the two cases. The arrows in the horizontal axis indicate the case with the lowest energy. \textbf{(b)}, \textbf{(d)} and \textbf{(f)}  show contour plots of the Kernel Density Estimation of the ground-state energy error of the quantum circuit optimized with orbital rotations $\Delta E_{\textrm{R}}$ versus the error of the same circuit optimized without orbital rotations in the MO basis $\Delta E_{\textrm{NR}}$. The kernel is an exponential kernel of bandwidth $1\textrm{mHa}$ The diagonal dashed line shows the points where $\Delta E_{\textrm{R}} = \Delta E_{\textrm{NR}}$ for reference. The vertical and horizontal dashed lines show the position of the error of the Hartree-Fock solution.}
\end{figure*}

\subsection{Variational Quantum Eigensolver with hardware-efficient circuits \label{Sec:VQE}}
The previous section showed a clear improvement in the variational energy when NQS in a VMC setting are dressed with parametrized and optimizable orbital rotations. In this section we study the improvements in the variational energy for a different class of variational states: hardware-efficient parametrized quantum circuits in a VQE setting. In this setting we not only investigate the improvement in the energy error, but also whether it is a consequence of the improvement in the variational power induced by the addition of orbital rotations or the change of the optimization landscape. 

In this case we map the spin-resolved modes of the system into individual qubits via the Jordan-Wigner mapping~\cite{JordanWigner1928}. Accordingly, a system with $N_{\textrm{orb}}$ orbitals is mapped into $2 N_{\textrm{orb}}$ qubits. 

The bare variational {\it ansatz} $|\psi_\theta\rangle$ is the hardware-efficient Cascade circuit introduced in Ref.~\cite{CunhaMotta2022VQEbenchmark}. This {\it ansatz} consists on the repeated and alternating application of single-qubit parametrized $R_y$ rotations followed the application of CNOT gates on adjacent qubits, on top of the $|00\hdots\rangle$ state of the computational basis. A single layer of this {\it ansatz} is shown in Fig.~\ref{fig_04:HF error} (a). This {\it ansatz} represents the identity operator if all $R_y$ rotation angles are set to $\theta = 0$. Furthermore, the action of Cascade layers can be turned off and on by setting the rotation parameter of the layer to zero or non-zero values respectively. As a consequence, given a circuit with $L$ cascade layers and optimized $R_y$ angles, it is possible to initialize a circuit with $L+1$ cascade layers with the parameters of the previous optimization and the parameters of the new layer set to small noise, allowing to incrementally improve the variational power of the {\it ansatz}.

The cascade {\it ansatz} is not number or magnetization preserving. Therefore, the optimization procedure is in charge of finding the state with the right number of spin-up and spin-down electrons. Furthermore, as the first and last qubits are not connected by CNOT gates, the orbitals associated to each qubit via the Jordan Wigner mapping cannot be all treated on equal footing.  Dressing the Cascade {\it ansatz} with with orbital rotations allows for an overall SWAP operation between qubits at the end of the circuit. The unitary $U(\kappa)$ is capable of not only constructing linear combinations of orbitals, but also permuting their indices. The addition of orbital rotations to the hardware-efficient {\it ansatz} produces a hybrid between a chemistry-inspired and hardware-efficient state.

While orbital rotations can be exactly described by a parametrized quantum circuit~\cite{Arute2020HFinQC}, they are implemented exactly as a classical post-processing step, leading to shallower quantum circuits and therefore less sources of noise. 

In this section we study the dissociation of the $\textrm{HF}$ molecule in the minimal STO-6G basis set. This system has strong dynamic (static) electronic correlations for small (medium to large) bond lengths $R$. The ground state for every bond length lives in the charge neutrality subspace with equal number of spin-up and spin-down electrons~\cite{CunhaMotta2022VQEbenchmark}. The VQE is emulated using the exact diagonalization tools of the software package QuSpin~\cite{Quspin2017}.

\paragraph*{\textbf{Energy error in the $\textrm{HF}$ molecule.--}} 
Figure~\ref{fig_04:HF error} (b) shows the variational ground-state energy obtained with the one-, two- and -three layer cascade {\it ansatz} dressed with orbital rotations. All three cases show very good agreement with the qualitative and quantitative nature of the dissociation curve of the molecule. Figure~\ref{fig_04:HF error} (c) shows the ground state energy error as a function of bond length $R$ for the Cascade {\it ansatz} with one, two and three layers. The error of the bare Cascade {\it ansatz} (optimized in the fixed MO basis) is compared to the error of the Cascade {\it ansatz} dressed with orbital rotations. As in the VMC case, the variational energies are significantly improved when variational orbital rotations are included. 

In the case of the Cascade {\it ansatz} with a single layer and when dynamical correlations are predominant (small $R$) the energy error is improved by several $\textrm{mHa}$, pushing the energy error below the $1\; \textrm{mHa}$ threshold for a significant number of the considered bond lengths. At moderate to large bond lengths the improvement in the variational energy is not as drastic, as the bare Cascade {\it ansatz} already achieves accurate variational energies. The same behaviour is observed for the Cascade {\it ansatz} with two and three layers. It must be noted that the dressed Cascade {\it ansatz} with three layers achieves energy errors below $1\; \textrm{mHa}$  for the entire dissociation curve. The variational energies of the bare {\it ansatz} with two and three layers shows a nonphysical cusp for bond length $R \sim 2.73\; \textrm{\AA} $. The energy cusp does not appear for the dressed counterparts, demonstrating that the results including variational basis optimizations are not only quantitatively more accurate, but they are also qualitatively correct.

The inset in the top panel of Fig.~\ref{fig_04:HF error} (c) compares the variational energy of the bare Cascade {\it ansatz} with three layers to the dressed Cascade {\it ansatz} with a single layer. It shows that the variational energies of both cases are very similar. In some cases, the dressed single-layer Cascade {\it ansatz} achieves better energies than the bare three-layer {\it ansatz}. These results demonstrate that at the expense of polynomial-scaling classical post-processing, the depth of variational quantum circuits can be significantly reduced (in this case by a factor of three) to obtain equivalent accuracies.

\paragraph*{\textbf{Energy optimization characteristics in the $\textrm{HF}$ molecule.--}}
In order to better understand the effect of the orbital rotations in the improvement of the variational energies, we analyze the optimization characteristics of the variational problem in both the presence and absence of orbital rotations.

For bond lengths $R = 1.21\; \textrm{\AA}$, $R = 1.46\; \textrm{\AA}$ and $R = 2.73\; \textrm{\AA}$ and for the Cascade {\it ans\"{a}tze} with one and two layers, 128 different circuit parameter initializations are considered. These parameter initializations are constructed by setting each $R_y$ rotation angle to a uniformly-sampled random value between $0$ and $2\pi$. Then, each circuit is optimized both with and without orbital optimizations. The single-particle basis of reference for the circuits optimized without orbital rotations is the MO basis.

Panels (a), (c) and (e) of Fig.~\ref{fig_04:HF statistics} show the cumulative percentage of randomly-initialized circuits that converge to a ground-state energy error $\Delta E$ or smaller, for the three different bond lengths under consideration. The energy error for the example with the smallest energy found is indicated by arrows. In all cases the lowest energy error was obtained by considering orbital optimizations. This indicates that the variational power of the dressed {\it ansatz} is indeed larger than the expressive power of the bare {\it ansatz}. 

We also observe that for all bond lengths and number of layers in the Cascade {\it ansatz}, the percentage of randomly initialized circuits that converge to energy error $\Delta E$ or smaller is significantly larger for the dressed variational states. For example, for $R = 1.21 \; \textrm{\AA}$ and the two-layer Cascade {\it ansatz}, $\sim 75 \%$ of the randomly initialized circuits dressed with orbital rotations obtained energy errors smaller than $10\; \textrm{mHa}$, while only $\sim 35 \%$ of the bare circuits converged to energy errors smaller than $10\; \textrm{mHa}$. 

We can therefore infer that the optimization landscape is less pathological when the single-particle basis is optimized together with the correlated variational {\it ansatz}. At a fixed single-particle basis, if the bare variational state is stuck at a local minimum or a region with small gradients, rotating the Hamiltonian to a different basis changes the optimization landscape, potentially placing the variational {\it ansatz} on a point of parameter space with non-vanishing gradients.

Panels (b), (d) and (f) of Fig.~\ref{fig_04:HF statistics} show the Kernel Density Estimation of the ground-state energy error of the quantum circuit optimized with orbital rotations $\Delta E_\textrm{R}$ versus the
error of the same circuit optimized without orbital rotations in the MO basis $\Delta E_\textrm{NR}$. A common feature shared amongst the bare Cascade circuits of different layers is that their tendency to get stuck at a point in parameter space corresponding to the the Hartree-Fock solution of the problem. Orbital rotations help those cases escape the point in parameter space that corresponds to that mean-field solution, yielding better variational energies. It must be noted that there is a small fraction of random initializations where the dressed quantum circuits also get stuck in points of parameter space that correspond to the mean-field solution. This fraction is much smaller than the corresponding fraction for the bare circuits.

 \section{Conclusions}
 In this manuscript we have shown that the addition of parametrized single-particle orbital rotations to different families of correlated variational wave functions, highly improves the accuracy of the achieved ground states in a wide range of physical systems. The improvement on the variational energy is due to two factors: the increase on the variational power of the ansatz and the improvement of the energy landscape in the space of variational parameters. The improvement of the energy landscape is a consequence of the treatment on equal footing the sets of the orbital rotation, and and the bare {\it ansatz} parameters in the optimization procedure. The orbital optimization adapts the basis that defines the Hamiltonian so that the expressive power of the variational wave function is maximal to represent the ground state. 
 We show numerical evidence demonstrating that this improvement in the accuracy persists, and even becomes more prevalent, for highly compact trial wave functions based NQS active space calculations.
 
 In the context of VMC it has the additional implication of allowing to variationaly adapt the single-particle basis that is used to specify the wave function amplitudes. In some cases, we also observe that the Hamiltonian can be transformed into a basis where the ground state has a simpler (and even positive semidefinite) nodal structure, or where the wave function amplitudes have a smooth structure in the space of Fock occupancies. This situation is favorable for non-determinant based  {\it ans\"{a}tze}.

 In the VQE framework, the orbital optimizations are treated as a classical post-processing step with polynomial scaling. The improvement of the expressive power of the variational circuits due to the classical post-processing allows to achieve comparable levels of accuracy with much shallower circuits. Shallower circuits are less susceptible to the intrinsic noise of quantum hardware. Therefore, this approach is greatly advantageous in NISQ-era devices. 

 Being agnostic to the class of variational state or variational algorithm, this approach can be applied to a wide range of scenarios, including the DMRG algorithm, where the orbital rotations would find the single-particle basis for which the ground state can be represented with a minimal amount of entanglement entropy.

\begin{table*}
    \centering
    \begin{tabular}{| c || c | c | c |}
    \hline
       Orbital group name  & \makecell{Occupancy \\ (for each spin-species)}  & \makecell{Excitations from/to \\ these orbitals}  & Involved in orbital rotations \\
       \hline\hline
        \makecell{Frozen \\ occupied} & 1 & No & No\\
        \hline
        \makecell{Inactive \\ occupied} & 1 & No & Yes\\
        \hline
        \makecell{Active \\ occupied} & \makecell{Active-space \\ wave-function dependent} & Yes & Yes\\
        \hline
        \makecell{Active \\ virtual} & \makecell{Active-space \\ wave-function dependent} & Yes & Yes\\
        \hline
        \makecell{Inactive \\ virtual} & 0  & No & Yes\\
        \hline
        \makecell{Frozen \\ virtual} & 0 & No & No\\
         \hline
    \end{tabular}
    \caption{Nomenclature and characteristics of orbitals in active-space calculations.}
    \label{tab:orbital notations}
\end{table*}

\begin{acknowledgements}
    The authors would like to thank Tomohiro Soejima, Miguel Morales, Ryan Levy, Shiwei Zhang and Antoine Georges and Antonio Mezzacapo for stimulating conversations. The Flatiron Institute is a division of the Simons Foundation. DS was supported by NSF: Grant OAC-2118310. 
\end{acknowledgements}
\appendix


\section{Active space energy evaluation}\label{Appendix: active space energy}
This appendix provides the expressions for the expectation values required to evaluate the energy of the system given an active space wave function.

Table~\ref{tab:orbital notations} provides a reference to the nomenclature to the different groups of orbitals involved in an active-space calculation. In our work we never consider frozen-occupied or frozen-virtual orbitals. All orbitals are involved in the basis optimization.

Consider the subset $\mathcal{A}$ of allowed active space electronic configurations defined by fixing the occupancy of a set of inactive spin-orbitals to be one and fixing the occupancy of a set of inactive virtual orbitals to be zero. The remaining electrons (active electrons) are allowed to have any configuration in the remaining set of active (occupied and virtual) orbitals. The highly compact active-space wave function is given by:
\begin{equation}
    |\psi_\theta \rangle = \sum_{n \in \mathcal{A}} \psi_\theta(n) |n \rangle.
\end{equation}
From now on, capital indices $PQRS$ are used to label inactive (occupied and virtual) orbitals while lower case indices $tuvw$ are used to label the active orbitals (occupied and virtual).   

The one-body reduced density matrix becomes very sparse and structured, taking the values:
\begin{equation}
\begin{split}
    \langle \psi_\theta| \sum_\sigma \hat{c}^\dagger_{P\sigma} \hat{c}_{t\sigma}| \psi_\theta\rangle & = 0 \textrm{ for all P, t}. \\
    \langle \psi_\theta| \sum_\sigma \hat{c}^\dagger_{P\sigma} \hat{c}_{Q\sigma}| \psi_\theta\rangle & = 2 G_{PQ} \\
    \langle \psi_\theta| \sum_\sigma \hat{c}^\dagger_{t\sigma} \hat{c}_{u\sigma}| \psi_\theta\rangle & = D_{tu}.
\end{split}
\end{equation}
$D_{tu}$ has a non-trivial expression and is defined by the equation above. The value of the $G_{PQ}$ matrix is given by:
\begin{equation}
    G_{PQ} = \begin{cases}
        \delta_{PQ} & \textrm{ if P and Q } \in  \textrm{ occupied}.\\
        0 & \textrm{ otherwise.}
    \end{cases}
\end{equation}
The non-zero two-body reduced density matrix elements can be separated into three groups. The first group is that where all of the indices belong to the inactive orbitals, whose matrix elements can be obtained directly from Wick's theorem:
\begin{equation}
    \langle  \psi_\theta| \sum_{\sigma\sigma'} \hat{c}^\dagger_{P\sigma}  \hat{c}^\dagger_{Q\sigma'} \hat{c}_{S \sigma'} \hat{c}_{R \sigma}| \psi_\theta \rangle = 4G_{PR} G_{QS} - 2 G_{PS} G_{QR}.
\end{equation}
The second group is the group for which all indices belong to the active orbitals. In this case there is no trivial expression for these elements:
\begin{equation}
    \langle  \psi_\theta| \sum_{\sigma\sigma'} \hat{c}^\dagger_{t\sigma}  \hat{c}^\dagger_{u\sigma'} \hat{c}_{w \sigma'} \hat{c}_{v \sigma}| \psi_\theta \rangle = D_{tuwv}.
\end{equation}
The third and last group corresponds to processes where an electron transitions between active space orbitals (from $v$ to $t$ with the same spin), while another electron transitions between inactive orbitals (from $S$ to $Q$ with the same spin):
\begin{equation}
    \begin{split}
        \langle  \psi_\theta| \sum_{\sigma\sigma'} \hat{c}^\dagger_{t\sigma}  \hat{c}^\dagger_{Q\sigma'} \hat{c}_{S \sigma'} \hat{c}_{v \sigma}| \psi_\theta \rangle & = 2 G_{QS} D_{tv}, \\
        \langle  \psi_\theta| \sum_{\sigma\sigma'} \hat{c}^\dagger_{Q\sigma}  \hat{c}^\dagger_{t\sigma'} \hat{c}_{v \sigma'} \hat{c}_{S \sigma}| \psi_\theta \rangle & = 2 G_{QS} D_{tv}, \\
        \langle  \psi_\theta| \sum_{\sigma\sigma'} \hat{c}^\dagger_{Q\sigma}  \hat{c}^\dagger_{t\sigma'} \hat{c}_{S \sigma'} \hat{c}_{v \sigma}| \psi_\theta \rangle & = - G_{QS} D_{tv}, \\
        \langle  \psi_\theta| \sum_{\sigma\sigma'} \hat{c}^\dagger_{t\sigma}  \hat{c}^\dagger_{Q\sigma'} \hat{c}_{v \sigma'} \hat{c}_{S \sigma}| \psi_\theta \rangle & = - G_{QS} D_{tv}.
    \end{split}
\end{equation}
Consequently, the expectation value of the Hamiltonian becomes:
\begin{equation}
\begin{split}
    \langle \psi_\theta | \hat{H} | \psi_\theta \rangle &  = \sum_{PQ} h_{PQ} \left(2G_{PQ} \right) + \sum_{tu} h_{tu} D_{tu}  \\
    & + \sum_{tvQS} D_{tv}G_{QS} \left(2 \frac{h_{tQSv}}{2} + 2 \frac{h_{QtvS}}{2} \right) \\
    & - \sum_{tvQS} D_{tv}G_{QS} \left( \frac{h_{QtSv}}{2} +  \frac{h_{tQvS}}{2} \right)\\
    & + \sum_{PQRS} \frac{h_{PQRS}}{2} \left(4G_{PR} G_{QS} - 2 G_{PS} G_{QR} \right)\\
    & + \sum_{tuvw} \frac{h_{tuvw}}{2} D_{tuwv}.
\end{split}
\end{equation}

\bibliographystyle{apsrev4-2}
\input{main.bbl}
\end{document}

%% file: main.bbl
\providecommand{\noopsort}[1]{}\providecommand{\singleletter}[1]{#1}%